\documentclass[aps,pre,showpacs,twocolumn,floatfix]{revtex4-1}
\usepackage{epsfig,graphicx,amsmath,amssymb}
\usepackage{bm}
\usepackage{color}
\usepackage{slashbox}
\usepackage{citesort}
\usepackage[normalem]{ulem}

\newcommand{\C}[1]{{\mathcal{#1}}}

\begin{document}

\title{Front structure and dynamics in dense colonies of motile bacteria: Role of active turbulence} 
\author{%
Rayan Chatterjee, Abhijeet A. Joshi,  and Prasad Perlekar}

%%%%%%%%% Insert author address here
\address{TIFR Centre for Interdisciplinary Sciences,
21 Brundavan Colony, Narsingi, Hyderabad.}
%%%% Keyword entries to be placed here %%%%
%\keywords{xxxx, xxxx, xxxx}

%%%% Abstract text to be placed here %%%%%%%%%%%%
\begin{abstract}
We study the spreading of a bacterial colony undergoing turbulent like collective motion. We present two minimalistic models to investigate the interplay 
between population growth and coherent structures arising from turbulence. Using Direct Numerical Simulation (DNS) of the proposed models we find that turbulence 
has two prominent effects on the spatial growth of the colony: (a) the front speed is enhanced, and (b) the front gets crumpled. Both these effects, which we highlight 
by using statistical tools,  are markedly different in our two models. We also show that the crumpled front structure and the  passive scalar fronts in random flows are related 
in certain regimes.
\end{abstract}
%%%%%%%%%%%%%%%%%%%%%%%%%%%

%%%%%%%%%% Insert the texts which can accomdate on firstpage in the tag "fmtext" %%%%%

%%%%%%%%%   End of first page  %%%%%%%%%%%%%%%

\maketitle

\section{Introduction}
Motile bacteria (e.g., \textit{Bacillus subtilis}) colonies form spectacular patterns as they spread on the surface of a Petri dish~\cite{wak94,jac94,ver08,swi09,kor10,den14}. The exact pattern depends on a variety of bio-physical conditions such as nutrient and agar concentration \cite{wak94}, motility \cite{jac94} etc. In a nutrient-rich environment, on a soft or hard agar plate, homogeneous spreading is observed. On a soft-agar plate, at low bacteria densities, spreading happens because bacteria perform run-and-tumble motion \cite{wak94}. On a hard agar plate, on the other hand, dense colonies of non-motile bacteria spread because individuals push each other as they reproduce \cite{kor10}.  

At moderate densities bacteria perform collective motion to form swarms \cite{kea10,wen}. Such swarming colonies form a variety of patterns, such as nearly homogeneous, concentric rings, and dendritic branches \cite{czi00,cop09,kea10}. More recent studies have revealed that at high concentrations, bacterial suspensions can show collective motion which strikingly resembles fluid turbulence~\cite{sok07,sok09,wen}.  The size and speed of typical collective structures is found to be an order of magnitude larger than the speed and size of a bacterium. Remarkably, similar to fluid turbulence the bacteria velocity field shows power-law correlations.  Not surprisingly, therefore, recent studies have used Navier-Stokes like equations to successfully model the velocity field of a turbulent bacterial suspension \cite{wen,bra15}.

Earlier numerical studies have modeled colony morphologies by using coupled reaction-diffusion type equations~\cite{gol98,lac99}. Homogeneously spreading colonies of non-motile bacteria  
have been successfully modeled using the Fisher equation,~Eq.~\eqref{eq:fish}.
\begin{equation}
\frac{\partial c}{\partial t} = D\nabla^2 c + \mu c \left(1-\frac{c}{Z}\right). 
\label{eq:fish}
\end{equation}
Here $c({\bm x},t)$ denotes the concentration of a bacterial colony,  $\mu$ is the reproduction rate, $D$ is the diffusivity that models the motion that arises because bacteria push each other as they grow and reproduce, and $Z$ is the carrying capacity that we set to $1$. Several studies have successfully used modified forms of Eq.~\eqref{eq:fish} to study growth of bacteria in different nutrient and agar conditions on a Petri dish. The Fisher equation and its variants have also been used to study competition between two species  \cite{gol98,swi09,kor10, swi09,pig13}.  Here, $c({\bm x},t)$ should be interpreted as the volume fraction of one of the two colonies. The Fisher equation coupled to Navier-Stokes equations has also been  used successfully to study coupling between hydrodynamics and chemistry \cite{bran,bha15}.

How does the collective motion of bacteria modify the spreading of a colony? For swarming vortex morphotype colonies \cite{ben98}, modeling the collective velocity field is essential to observe the correct spreading pattern~\cite{czi96,ben97}. However, to the best of our knowledge, there are still no experimental studies on the growth of colonies in the recently found regime of bacterial turbulence. In this paper we undertake an exploratory study to investigate the role of turbulent-like collective motion on colony spreading. Following the classical work of Fisher~\cite{fis37},  we assume an abundance of nutrients and a homogeneous environment.

We present two minimalistic models to numerically investigate the spreading of a dense bacterial suspension that performs turbulent-like collective behavior.  Our study shows that the collective motion: (a) speeds up the spreading of a colony and (b) the colony front gets crumpled as it propagates. The crumpling at the frontiers is qualitatively similar to the plankton patterns on the ocean surface, the difference being that in dense bacterial suspensions, stirring is internal whereas, background flow advects plankton \cite{abr98,mar03,per10}.

The rest of the paper is organized as follows. We first introduce the models that we use to study the spreading of a colony. Next we give an overview of the numerical method that we use. We then discuss the results obtained from our numerical simulations.  We conclude by providing a discussion of our results. 

\section{Model}
Motivated by Wensink \textit{et al.} \cite{wen}, we model the motion of a turbulent bacterial colony using the following equation for the velocity field. As we are interested in dense bacterial colonies, we assume density variation is negligible and 
enforce an incompressibility constraint $\nabla \cdot {\bm v}=0$ \cite{wen}:
\begin{eqnarray}
\frac{\partial {\bm v}}{dt}&=& \lambda{\bm v}\times {\bm \omega} -\nabla p + (\alpha(c)-\beta |v|^2){\bm v}\nonumber\\
                                  &&+ \Gamma(c) \nabla^2 {\bm v}-\Gamma_2\nabla^4 {\bm v}
\label{eq:act}.
\end{eqnarray}
Here ${\bm v}({\bm x},t)$, $\omega({\bm x},t)$, and $p({\bm x},t)$ are continuous fields that describe the velocity, the vorticity, and the pressure field of a dense bacterial suspension, 
the coefficients $[\Gamma(c),\Gamma_2]$ are the strength of the small-scale stirring and damping, and the coefficient $\lambda$ of the Navier-Stokes-like term ${\bm v}\cdot \nabla {\bm v}$ is 
in general non unity because of the lack of Galilean invariance \cite{ramaswamy1}.  The velocity magnitude $|{\bm v}|=\sqrt{\alpha(c)/\beta}$ in absence of all the gradient terms in Eq.~\eqref{eq:act}. 
$|{\bm v}|=0$ for $[\alpha(c)\leq0,\beta>0]$ and $|{\bm v}| >0$ otherwise. 
Because of the collective motion, the bacterial suspension also gets advected by the velocity field ${\bm v}$. This is easily modeled by supplementing Eq.~\eqref{eq:fish} with an advection term. The modified equation for the evolution of the concentration field is 
\begin{equation}
\frac{\partial c}{\partial t} + {\bm v}\cdot \nabla c = D\nabla^2 c + \mu c(1- c).
\label{eq:fishact}
\end{equation}
The equations that we use fall broadly under the Toner-Tu-Ramaswamy class of hydrodynamic equations for soft-active matter \cite{ber06,mis10,ramaswamy1,mar13,ton14,yan14,doo16}. The coefficients $\alpha(c)$ and 
$\Gamma(c)$ model the effect of bacterial concentration on the collective motion. As we are interested in planar growth of a colony on a Petri dish-like surface, we study dynamics in two dimensions. 

Below we present two possible choices of $[\Gamma(c)$,$\alpha(c)]$ which are of experimental relevance.
\begin{enumerate}
\item Model A, $\Gamma(c)\equiv\Gamma$ and $\alpha(c)\equiv\alpha$. We use model A to study the invasion of one bacterial colony into another. We assume that both the colonies have indistinguishable swimming capabilities and are in turbulent phase. For this model, it is more appropriate to think of  $c$ as the concentration of the invading species. 
\item Model B, $\Gamma(c)\equiv \Gamma c$ and $\alpha(c)\equiv \alpha c$.  We use model B to study spreading of a bacterial colony on a surface. Our choice $\alpha(c)=\alpha c$ and $\Gamma(c)=\Gamma c$ ensures that ${\bm v}=0$ when $c=0$.
\end{enumerate}
\section{Direct Numerical Simulations}
We use a square domain ${\mathcal D}$ with each side of length $L=32\pi$ and discretize it using $N^2=2048^2$ collocation points. We numerically integrate Eq.~\eqref{eq:fishact} using a second-order explicit finite-difference scheme for spatial derivatives and the Euler method for time integration \cite{per10}. To ensure incompressibility, we write Eq.~\eqref{eq:act} in vorticity-stream function formulation  Eq. \eqref{eq:actsv} and numerically integrate it using using a pseudospectral method \cite{per11}:
%\begin{equation}
\begin{eqnarray}
\frac{\partial \omega}{\partial t} &=& \lambda \nabla \times ({\bm v}\times \omega) + \nabla\times [\alpha(c) - \beta |v|^2]{\bm v} \nonumber\\
&&+ \nabla \times [\Gamma(c) \nabla^2 {\bm v}]  + \Gamma_2 \nabla^4 \omega. 
\label{eq:actsv}
\end{eqnarray}
%\end{equation}
Here, $\psi({\bm x},t)$ is the streamfunction, ${\bm v}= \hat{z}\times \nabla \psi$, and $\nabla^2\psi=\omega$.

We set $\alpha=1$, $\beta=0.5$, $\Gamma=-0.045$, $\Gamma_2=|\Gamma|^3$, and $\lambda=3.5$ so that the velocity correlation statistics are consistent with that of a quasi-2D {\it B. subtilis} suspension~\cite{wen}.

\begin{figure}[h]
\includegraphics[width=\linewidth]{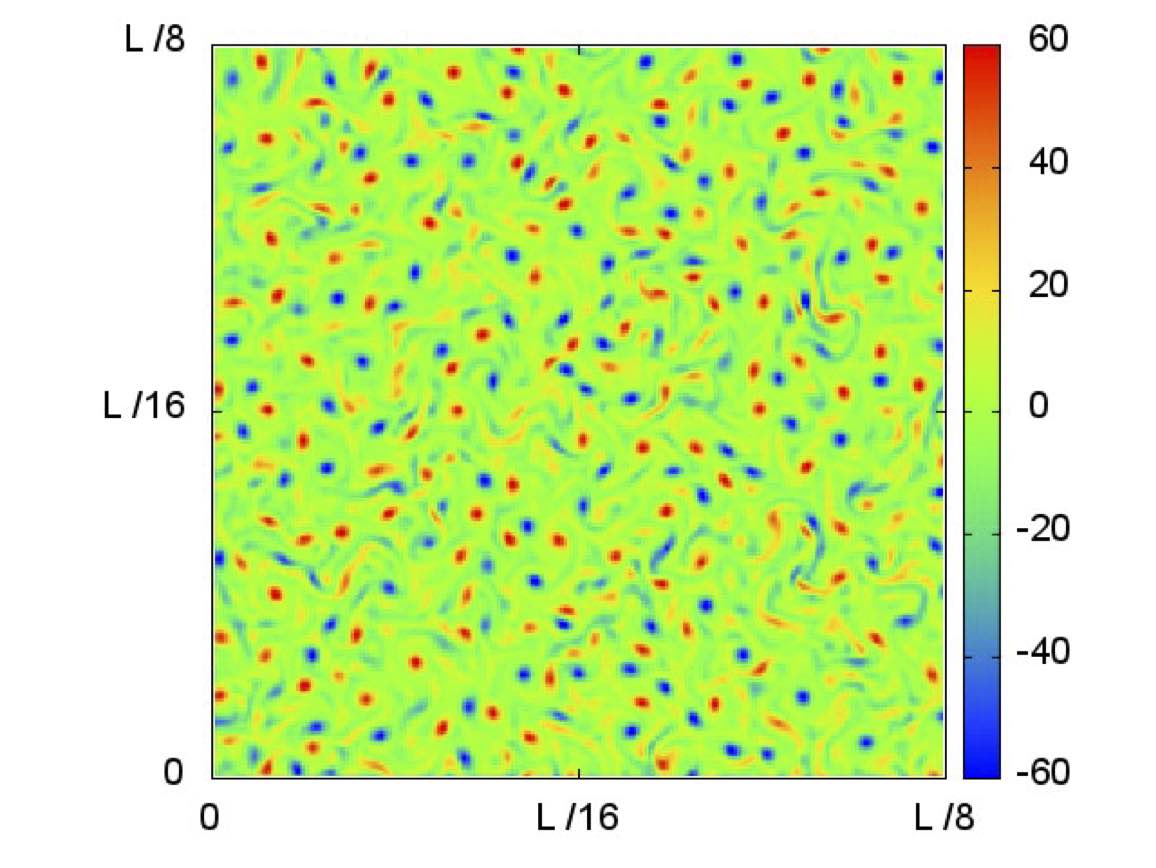}\\
\includegraphics[width=\linewidth]{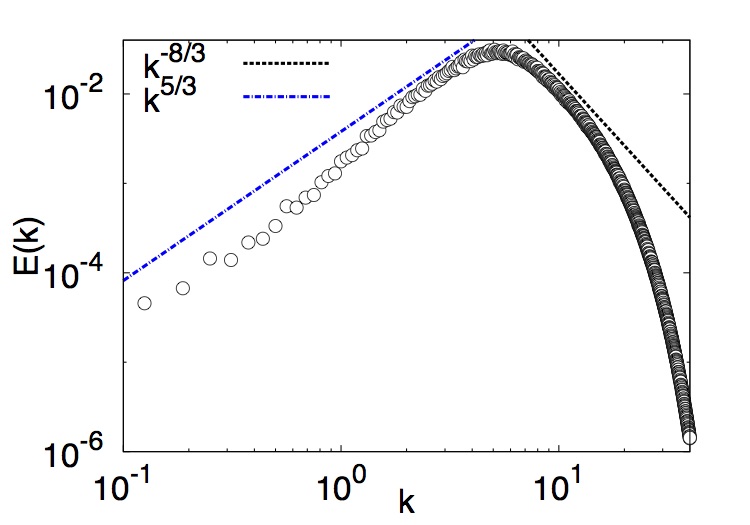}
\caption{\label{fig:fig1} (Top) The pseudocolor plot of the steady-state vorticity field over a section of our simulation domain obtained from  DNS of Eq.~\eqref{eq:act} with $\alpha(c)=\alpha$ and $\Gamma(c)=\Gamma$. (Bottom) The corresponding kinetic energy spectrum $E(k)=\sum_{k'} |u_{k'}^2|$ where $k' \in [k-1/2,k+1/2]$.  The peak of the spectrum occurs around $k_m=6$. In agreement with previous studies, we observe a $k^{5/3}$ scaling (blue dash-dot) for $k < k_m$ and a $k^{-8/3}$ scaling (black dash) for $k > k_m$.}
\end{figure}

In Fig.~\ref{fig:fig1} we show a typical snapshot of the vorticity field and the corresponding energy spectrum obtained by direct numerical simulation of Eq.~\eqref{eq:act} with $\Gamma(c)\equiv\Gamma$ and $\alpha(c)\equiv\alpha$. Note that the exponents of $5/3$ for low wave numbers and $-8/3$ for high wave numbers are consistent with Ref.~\cite{wen}. 

We initialize $c$ as 
\[
    c(x,y,t=0)= 
\begin{cases}
    1,& \text{if } x\leq L/100 \\
    0,              & \text{otherwise}
\end{cases}
\]
and study its evolution for varying diffusivity $D$ and growth rate $\mu$.

\section{Results}
In the absence of the velocity field ${\bm v}$, the concentration front of width $\sim 8\sqrt{D/\mu}$ propagates from left to right with a 
speed $\sim 2\sqrt{D \mu}$ (Fisher velocity) \cite{murray,fis37,kol37}. What happens when bacteria perform 
collective motion that resembles turbulence?

Using model A and model B, we now systematically characterize the properties of colonies performing 
turbulence like collective motion. We study how bacterial turbulence modifies the spatiotemporal structure 
of the spreading or invasion of a colony. We conduct measurements in the spatiotemporal window where the 
front moves with a constant velocity and is $L/3$ distance away from the left and right boundaries.

\begin{figure*}[!t]
    \centering
\includegraphics[width=\linewidth]{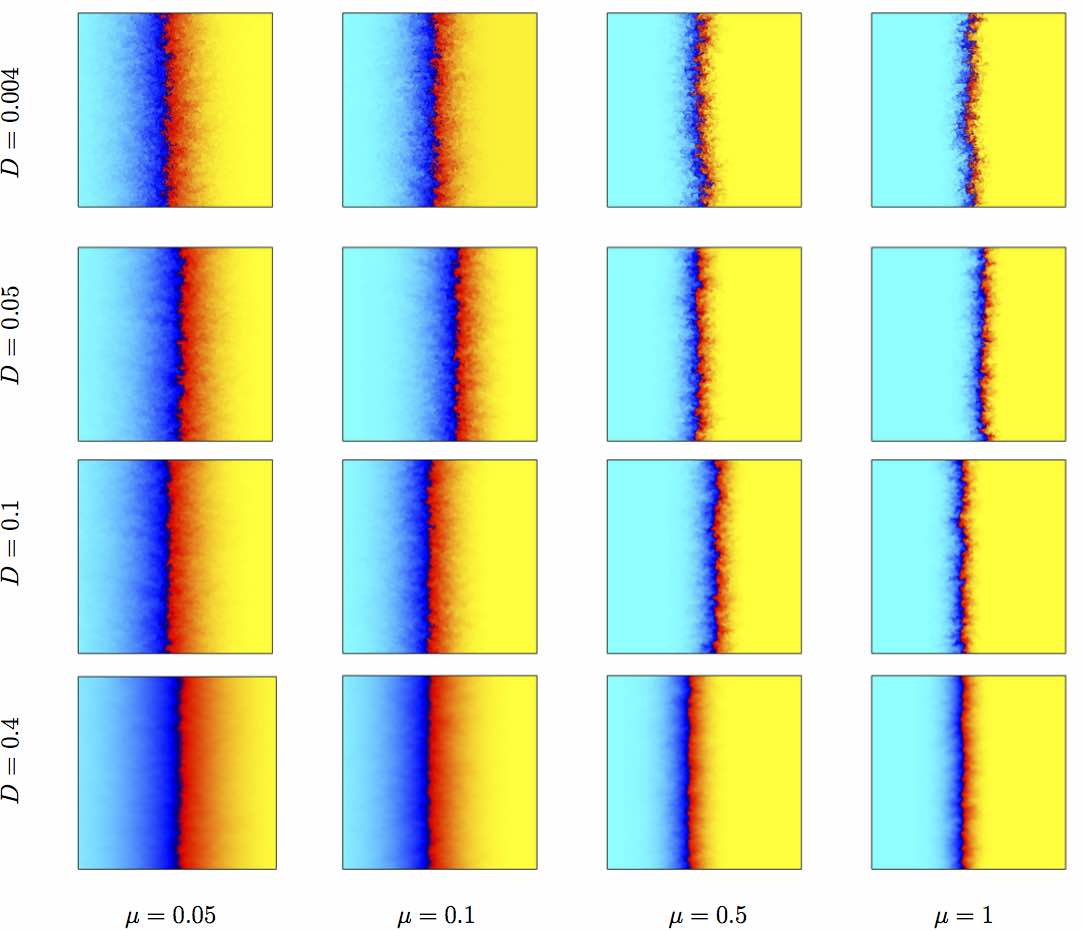}
\caption{\label{fig:domain_modelA} Pseudocolor plots of the concentration fields for model A for representative values of $D$ and $\mu$.
  To emphasize the front structure, we show a square window of side length $\approx L/4$. Blue indicates regions of high concentration ($c \geq 0.5$), and yellow indicates regions of low concentration.}
\end{figure*}

\subsection{Front propagation in Model A}
Using model A, we investigate the invasion of a motile colony with a selective advantage $\mu$ into 
another motile colony.  In Fig.~\ref{fig:domain_modelA} we show typical snapshots of the concentration profile for the representative values of diffusivity $D$ and growth rate $\mu$. The interface becomes rough because of the advecting velocity field ${\bm v}$. The interface roughness increases on reducing $D$ and $\mu$. In particular for a fixed $D$, the undulations of the concentration front become more compact on increasing $\mu$. On the other hand, for a fixed $\mu$, undulations of concentration front are enhanced on reducing $D$. Physically, a large value of $D$ implies that the motion because of bacteria pushing each other overwhelms the collective behavior. In this regime, as observed in Fig.~\ref{fig:domain_modelA}, we indeed find that collective motion has a very minor effect on the front. We quantify these observations in the following sections.

\subsection{Front propagation in Model B}
\begin{figure*}[!t]
\includegraphics[width=\linewidth]{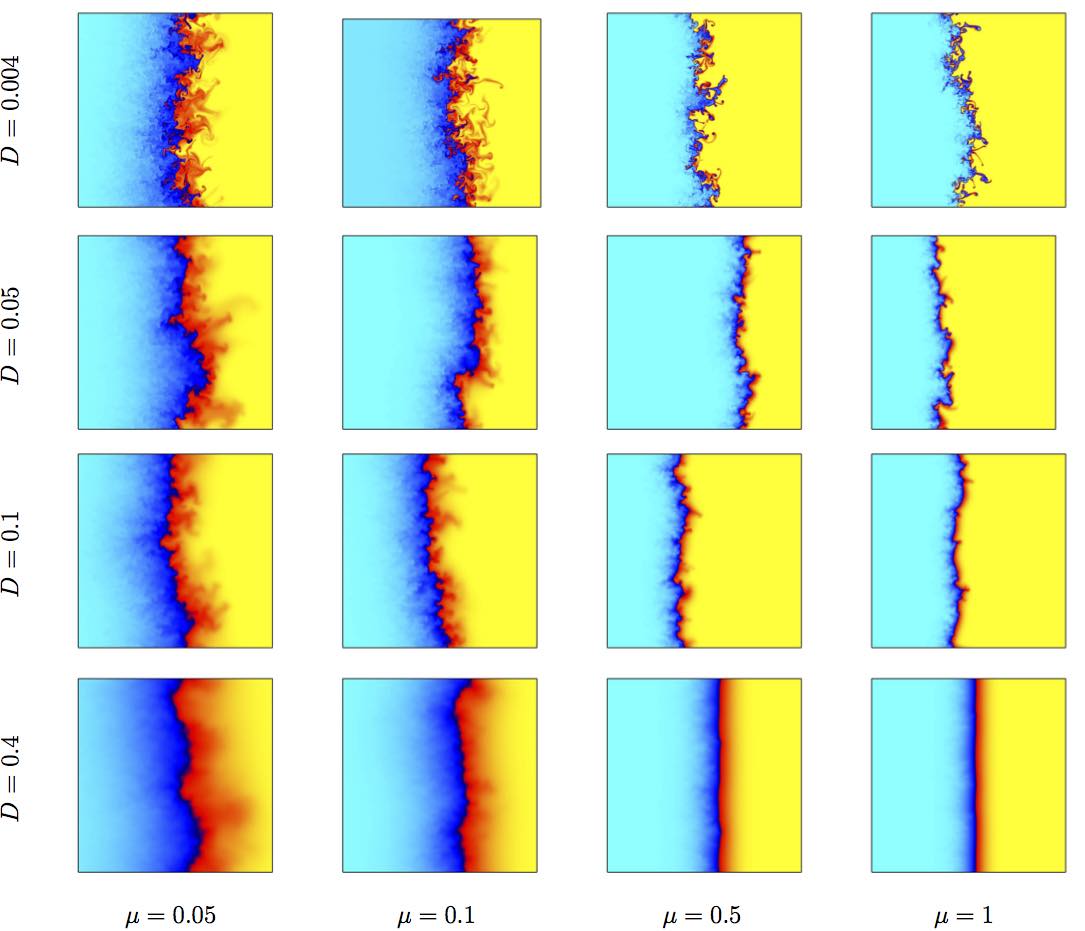}
\caption{\label{fig:domain_modelB} Pseudocolor plots of the concentration fields for model B for representative values of $D$ and $\mu$. We 
only show a square window with each side $\approx L/4$ to emphasise the front-structure. Blue indicates regions of high concentration ($c \geq 0.5$) and yellow indicates regions of low concentration. Note that interface undulations in model B are larger in comparison to model A (see Fig.~\ref{fig:domain_modelA}).} 
\end{figure*}

We use model B to investigate spreading of a motile colony with doubling time $\mu$. In Fig.~\ref{fig:domain_modelB} we show typical snapshots of the concentration field for the representative values of diffusivity $D$ and growth rate $\mu$. Here again, the presence of collective motion leads to roughing of the interface.  However, unlike model A, in model B velocity is present only where bacteria concentration is nonzero. This leads to formation of finger-like patterns in model B that are absent in model A for the same values of $D$ and $\mu$ (compare Figs. \ref{fig:domain_modelA} and \ref{fig:domain_modelB}).

\subsection{Front speed: Model A versus Model B \label{sec4c}}
\begin{figure}[!h]
 \includegraphics[width=\linewidth]{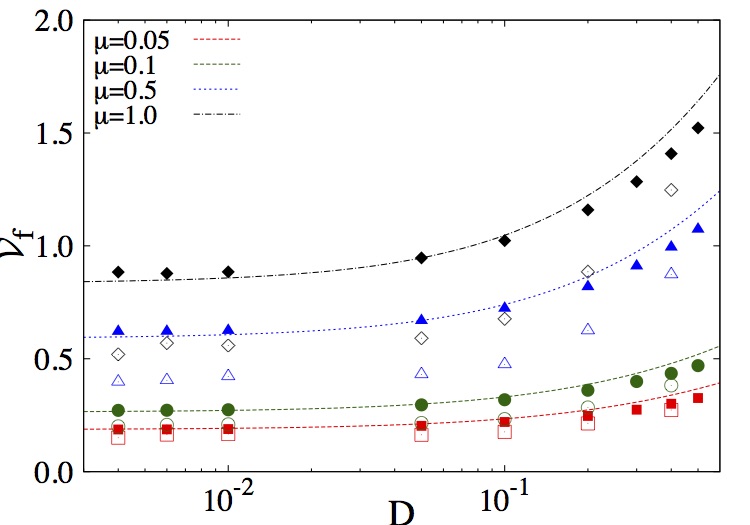}
 \caption{ \label{fig:speed2} Turbulent front speed  $\C{V}_f$ versus diffusivity  $D$ for $\mu=0.05$ (square), $\mu=0.5$ (circle), $\mu=0.1$ (triangle), and $\mu=1.0$ (diamond) for model A (filled symbols) and model B (empty symbols). Dashed lines show the corresponding front speed estimated by eddy diffusivity approximation $\C{V}_f=2\sqrt{(D+D_t) \mu}$ with $D_t=0.17$ for model A.}
\end{figure}

We now investigate the speed of the concentration front for the two models. The front speed is calculated as
\begin{eqnarray}
\C{V}_f = \frac{d}{dt} \left [\frac{1}{L}\int c({\bm x},t) dx dy \right ].
\end{eqnarray}
We have verified that in absence of ${\bm v}$, $\C{V}_f=2\sqrt{D \mu}$. As turbulence enhances the effective diffusivity 
of a scalar (e.g., temperature), in the same way we expect that presence of motility (bacterial turbulence) would enhance bacterial 
diffusivity $D$ and hence $\C{V}_f$. In Fig.~\ref{fig:speed2} we plot $\C{V}_f$ versus $D$ for 
the two models for $\mu=0.05,0.1,0.5,$ and $1$.   It is clear that the front speed for model A is larger than model B. This is because 
for model A both the species are motile and hence ${\bm v}$ is non zero and of the same magnitude everywhere, whereas for model B, 
${\bm v}$ is non zero only where $c\neq 0$.  For model A, we can estimate the turbulent diffusivity as $D_{t}=v_0/k_m \approx 0.17$ where, $v_0\equiv\sqrt{\Gamma^3/\Gamma_2}=1$ is the characteristic 
velocity of the turbulent flow \cite{wen} (see Appendix for a detailed calculation of $D_t$).  Thus the front speed in the presence of turbulence for model A can be estimated as $2\sqrt{\mu (D+D_t)}$, which is in close agreement with the result of our DNS (see Fig. ~\ref{fig:speed2}).  In the limit $D\to0$, the front speed is completely determined by turbulent diffusivity $\C{V}_f\sim 2\sqrt{D_t \mu}$. This  explains the roughness of the interface at lower values of $D$. On the other hand, when $D \gg D_t$, collective motion is irrelevant and $\C{V}_f\sim2\sqrt{D \mu}$ for the two models.  

\subsection{Multivalued nature of the propagation front}

\begin{figure}[!h]
\includegraphics[width=\linewidth]{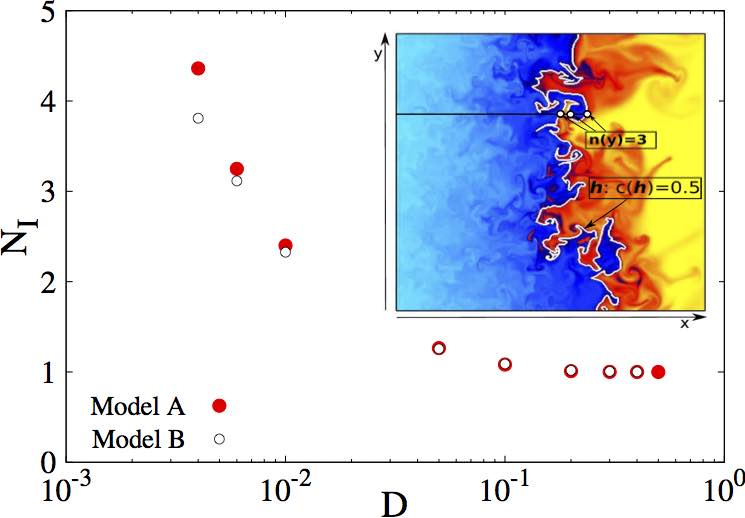}
\caption{ \label{fig:intersections} Average number of intersections ($N_I$) as a function of the diffusivity $D$ for $\mu=0.05$. For $D > 0.1$ the front is essentially single valued. 
We do not observe any significant dependence of $N_I$  on $\mu$ (not shown here). The inset shows a zoomed in snapshot of the concentration field along with the $c=0.5$ hull ${\bm h}$ (white curve) obtained by using BRWA for $D=0.004$ and $\mu=0.05$. Because of the underlying turbulence, the hull ${\bm h}$ is multivalued at several locations.}
\end{figure}

As a result of underlying turbulence, the front structure gets distorted. From Figs.~\ref{fig:domain_modelA} and \ref{fig:domain_modelB} it is clear that at the interface, $c({\bm x})\approx 0.5$. We define $N_I\equiv \langle \sum_{(i,j)} \delta[c(x_i,y_j,t)-0.5]/N\rangle$  as a preliminary estimator of the front structure. Here $0\leq(i,j)<N$ are the Cartesian grid indices in our simulation domain ${\mathcal D}$, and $0\leq$$\langle\cdot\rangle$ indicates temporal averaging. Thus for a front without overhangs, $N_I=1$, whereas $N_I=N$ if $c=0.5$ over the entire domain. The plot in Fig.~\ref{fig:intersections} shows that at large values of $D$, $N_I=1$, indicating the smooth nature of the front. On reducing $D$, $N_I$  keeps on increasing monotonically, indicating the enhanced roughness of the front. We do not observe any significant difference in $N_I$ between model A and Model B except for very small value of $D$. This qualitative dependence does not change on varying $\mu$. We would like to point that for small values of $D$, $N_I$ is larger for model A in comparison to model B. This is because in Model A turbulence is present over the entire domain and leads to enhanced stirring and formation of small-scale structures. The enhanced small-scale structure is also consistent with our earlier observations about larger front speeds $\C{V}_f$ for model A in comparison to model B (Sec. ~\ref{sec4c}, Fig.~\ref{fig:speed2}).

From visual inspection (see Figs.~\ref{fig:domain_modelA} and \ref{fig:domain_modelB}) it is clear that although model A and  model B have similar values of $N_I$, the sizes of interface undulations are significantly different for the two models (see Figs.~\ref{fig:domain_modelA} and \ref{fig:domain_modelB}). To quantify these differences, we first need to identify a front in the  concentration field $c({\bm x},t_0)$ at a time instant $t_0$.  We use the biased random walk algorithm (BRWA) \cite{gast1} to identify a locus of points (or a hull) ${\bm h}_i\equiv(x_i,y_i)$ such that $c({\bm h}_i,t_0)=0.5$, where the hull index $0\leq i \leq N_h$ and $0\leq (x_i,y_i) \leq L$ are the Cartesian points in our simulation domain ${\mathcal D}$. Connecting the points of the hull, we get a continuous curve that starts at the bottom of the domain $y=0$ and ends at the top $y=L$. Figure~\ref{fig:intersections}(inset) shows a representative plot of the $c=0.5$ hull overlaid on the pseudocolor plot of the concentration field.

\subsubsection{Hull width}
\begin{figure}[!h]
\includegraphics[width=\linewidth]{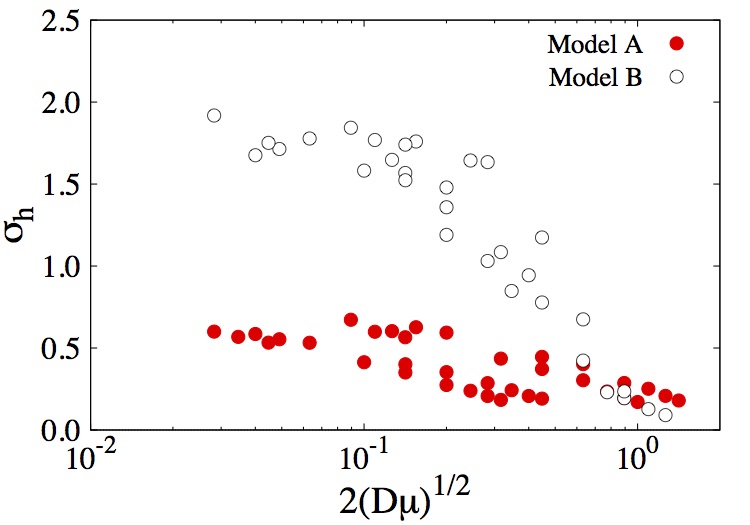}
 \caption{\label{fig:sigma_h} Standard deviation $\sigma_h$ of the front height with respect to its mean position as function of $2\sqrt{D\mu}$. Note that for 
 $v_0 \ll 2\sqrt{D \mu}$, $\sigma_h$ is dramatically different  for the two models indicating presence of large plume like structures in Model B.}
\end{figure}

We start our analysis by calculating the hull width $\sigma_h =\langle [\frac{1}{N_h}\sum_{i=0}^{N_h} x_i^2- (\frac{1}{N_h} \sum_{i=0}^{N_h} x_i)^2 ]^{1/2}\rangle$ (standard deviation of the $x$ coordinate of the hull).  Here, $\langle [\cdot]\rangle$ indicates temporal averaging. In Fig. ~\ref{fig:sigma_h}, we plot $\sigma_h$ as a function of $2\sqrt{D \mu}$ (the intrinsic front velocity in absence of collective motion) for the two models. When the typical turbulent velocity $v_0 \ll 2\sqrt{D \mu}$, the intrinsic diffusion dominates over turbulence and the two models behave in the same way. On the other hand for $v_0 \gg 2\sqrt{D \mu}$, $\sigma_h$ for model B is larger than model A. This is consistent with our observation about the presence of large, plume-like structures in model B (see Figs.~\ref{fig:domain_modelA} and \ref{fig:domain_modelB}).

\subsubsection{Hull fractal dimension}

\begin{figure*}
  \includegraphics[width=0.32\linewidth]{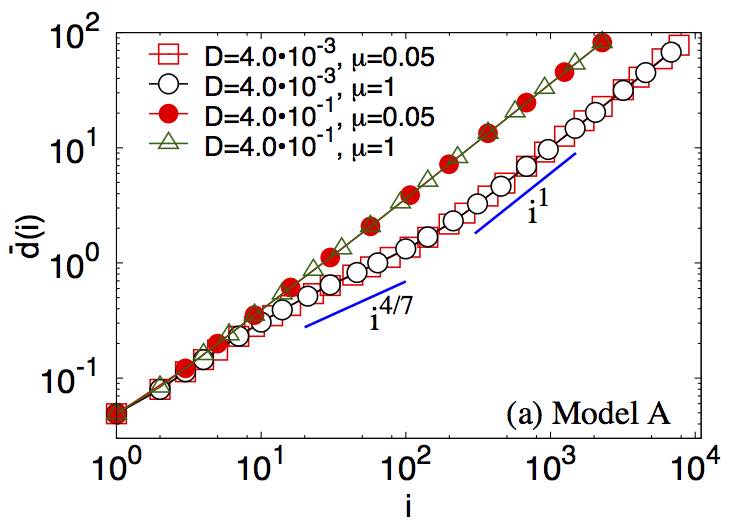} 
  \includegraphics[width=0.32\linewidth]{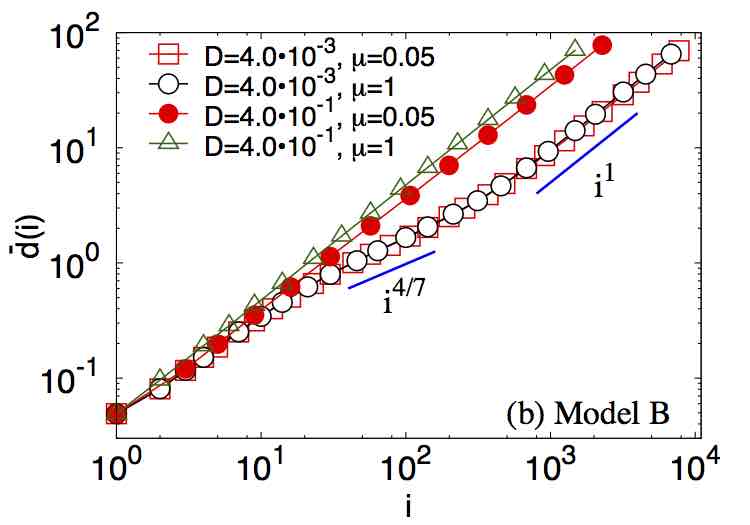}
  \includegraphics[width=0.32\linewidth]{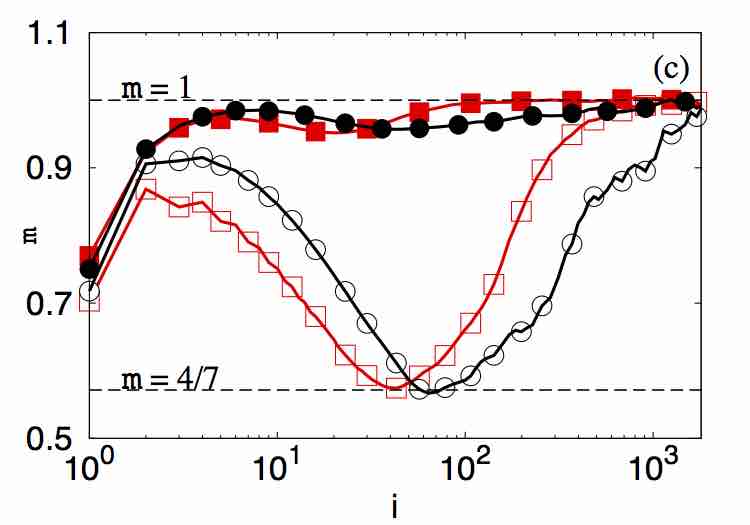}
   \caption{\label{fig:frac}Average distance between points $\overline{d}(i)$ versus distance index $i$ plotted on log-log axes for (a) model A and (b) model B for different values of $D$, and $\mu$. (c) Semilog plot of the local slope $m$ for model A [$D=4\cdot 10^{-3}$ (red empty square), and $D=4\cdot 10^{-1}$ (red filled square)] and model B [$D=4\cdot 10^{-3}$ (black empty circle), and $D=4\cdot 10^{-1}$ (black filled circle)] at fixed $\mu=5\cdot 10^{-2}$. Horizontal dashed lines indicate $m=4/7$ and $m=1$.}
\end{figure*}

 We now study the fractal dimension of the hull using an equispaced polygon method \cite{gast1}. Consider a hull consisting of a sequence of points $(x_0,y_0), (x_1,y_1).....(x_{N_h},y_{N_h})$, and the average distance 
 between points separated by $i$ steps is
  \begin{equation}
  \overline{d}(i)=\sum_{j=0}^{N_h-i} d_j(i)/(N_h-i+1).
  \end{equation}
   Here, $0\leq i \leq N_h-1$ and $d_j(i)=\sqrt{(x_j-x_{j+i})^2+(y_j-y_{j+i})^2}$. For a fixed number 
   of steps $i$, the average distance and the fractal dimension are related as $\overline{d}(i)\propto i^{1/d_f}$ \cite{gast1}. 
   In Figs.~\ref{fig:frac}(a) and ~\ref{fig:frac}(b) we plot $\overline{d}(i)$ versus $i$ for different values of $D$ and $\mu$ for the two models. For large 
   values of $D$, independent of $\mu$ and except for very small scales, we find that $\overline{d}\propto i$ i.e., the front is essentially flat $d_f=1$. For small values of $D$, the presence 
   of bacterial stirring leads to front undulations. We find a decade-long scaling range with $\overline{d}(i)\propto i^{4/7}$ or $d_f=7/4$ around the typical eddy scale ($\overline{d} \approx 2\pi/k_m$)  and $\overline{d}(i)\propto i$ for $i \gg 2\pi/k_m$. Note that $d_f=7/4$ also for purely diffusive fronts \cite{sap85}. Thus, $d_f=7/4$ further supports our modeling of bacterial turbulence by an effective diffusivity. To highlight the difference between model A and model B, in Fig.~\ref{fig:frac}(c) we plot the local slope $m\equiv {\rm d} \log{\overline d}/{\rm d} \log{i}$ versus $i$. As discussed earlier, we find that for large $D$, $m \to 1$. However, for small $D$ we observe that the region with $4/7$ scaling for model A appears at a slightly earlier stage than model B. We believe this is because in model A the bacterial stirring is present on both sides of the front, whereas for model B it is only present in regions with $c=1$. Similar cross over from $d_f \simeq 7/4$ to $d_f=1$ has also been observed in earlier studies  on front propagation in 2d microscopic simulations of diffusing particles \cite{sap85}, the stochastic Fisher-Kolmogorov-Petrovsky-Piskunov (sFKPP) equation \cite{lem}, and in vegetation fronts \cite{gast1}.   We would like to point out that in the case of sFKPP, the front undulations are driven by a stochastic noise that models fluctuations in the size of the bacteria population \cite{kor10,pig13}, whereas in our study collective motion of the bacteria causes front undulations and also sets up the scale at which cross over in $d_f$ takes place.

\section{Concentration Spectrum}
\begin{figure*}
\includegraphics[width=0.48\linewidth]{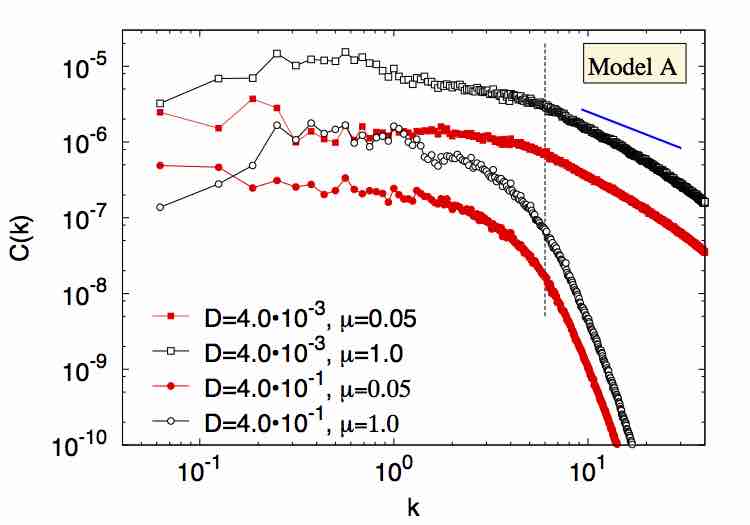} 
\includegraphics[width=0.48\linewidth]{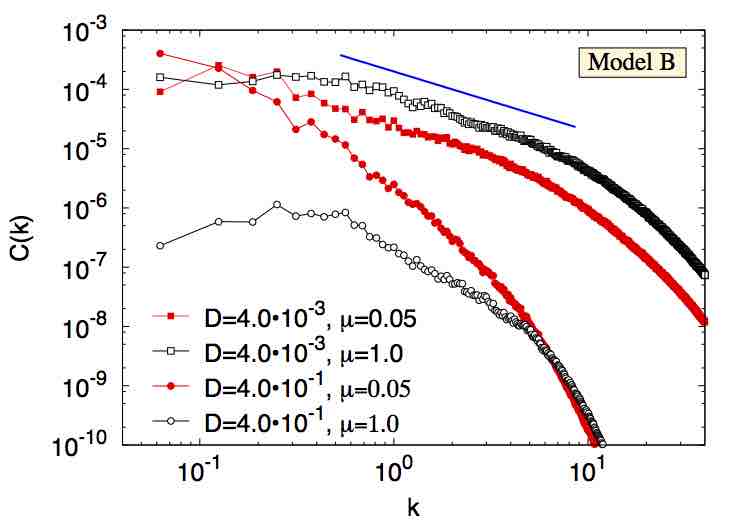}
 \caption{\label{fig:cspec}Concentration spectra for model A (left) and model B (right) for varying diffusivity $D=0.004$, and $0.4$, and $\mu=0.05$, and $1.0$.  The blue line indicates the Bachelor scaling $k^{-1}$ and the vertical dashed line indicates $k_m$.}
\end{figure*}
To further quantify the statistical properties of the undulating interface, we now study the spectrum of fluctuations in the concentration field arising from bacterial turbulence. This is expressed as : $C(k)=\sum_{k-1/2 \leq k' \leq k+1/2} |c_k'^2|$ where, $c^\prime = c - (\int c dy)/L$. The plot in Fig.~\ref{fig:cspec} shows $C(k)$ versus $k$ for model A and model B. \\

\medskip
{\it{$C(k)$ for model A}}  [Fig. ~\ref{fig:cspec}(left)].  The spectrum is flat and does not show any scaling behavior for $k<k_m$. For $k>k_m$ and large $D=0.4$, diffusion is dominant and the spectrum falls off sharply.  At small $D=0.004$, interface modulation because of turbulence becomes dominant and we observe a small regime showing $C(k) \sim k^{-1}$ scaling. The $k^{-1}$ scaling appears because at small scales the undulations because of stirring are similar to those of a passive-scalar stirred by random flow which shows the $k^{-1}$ Bachelor scaling  \cite{bat59,kra68}. \\

{\it{$C(k)$ for model B}}  [Fig. ~\ref{fig:cspec}(right)]. Here the spectral properties are more intriguing. For $D=0.4$ and $\mu=1.0$,  the amplitude variations are of the same order as model A, but we observe a $k^{-1}$ regime for $k<k_m$. For small $D$ and $\mu=0.05,1.0$, we observe that both large and small scale undulations are present (see Fig.~\ref{fig:domain_modelB}). This shows up as an extended $k^{-1}$ scaling regime in the Fourier space. The intermediate case with $D=0.4$ and $\mu=0.05$ is the most intriguing. We observe the presence of large-scale undulations but no small-scale plume-like structures or finger-like patterns  (Fig.\ref{fig:domain_modelB}). The $C(k)$ spectrum for this case is much steeper than $k^{-1}$, and the spectral content is close to the $D=0.004,\mu=0.05$ case for $k<k_m$ and is close to the $D=0.4,\mu=1.0$ case  for $k>k_m$.

\section{Conclusion}
We proposed two minimalistic models to study colony front propagation in dense colonies of motile bacteria performing turbulence like collective motion. We study two 
scenarios: (a) invasion of one colony over the other  (model A) and (b) spreading of colony on a Petri dish (model B). We find that the presence of collective turbulence-like motion always enhances the front propagation speed. We highlight the similarities and the differences between the two models.In particular, model B allows for large-scale undulations which are absent in model A. We quantify the fractal structure of the front and show that the fractal dimension of the front around the stirring scales is $d_f=7/4$. Finally, we also show that, for certain parameter values, the concentration fluctuations arising from bacterial turbulence are similar to those of passive scalar stirred by a random flow. Earlier experiments have investigated spreading of dense colonies of non-motile bacteria or of motile bacteria that form swarms.  We hope that our simulations will stimulate new experimental studies on the spreading of colonies in this regime of bacterial turbulence.

\section{Acknowledgments}%
We thank S. Ramaswamy, R. Govindarajan, and S. Shenoy for discussions and the Department of Atomic Energy (DAE), India, for financial support.

\appendix*
\label{app}

\section{Eddy diffusivity for model A}
\begin{table}[!t]
   \begin{tabular}{@{\extracolsep{\fill}}| c ||c | c | c | c |}
    \hline
     \backslashbox[6mm]{$D$}{$\mu$} & 0.05 & 0.1 & 0.5 & 1.0   \\
 \hline
 \hline
 0.004 & 0.20 & 0.21 & 0.18 & 0.18 \\ 
 \hline
 0.05 & 0.19 & 0.19 & 0.19 & 0.18 \\ 
 \hline
 0.1 & 0.17 & 0.18  & 0.18 & 0.16 \\ 
 \hline
 0.4 & 0.10 & 0.10 & 0.11 & 0.11 \\
 \hline
\end{tabular}
 \caption{\label{tab}Numerical estimate of turbulent diffusivity $D_t \equiv \overline{v_x c'}/\partial_x \overline{c}$ for different values of $D$ and $\mu$ obtained from 
 our direct numerical simulations.}
\end{table}
Using the procedure outlined in Ref.~\cite{bran}, we now briefly describe the methodology to obtain the eddy diffusivity for model A.  Assuming very small variations of 
the concentration field perpendicular to the direction of front propagation, we decompose these as 
 $c(x,y) = \overline{c}(x) + c'(x,y) $. Here the overline indicates averaging over the $y$ direction $\overline{f}(x)\equiv \frac{1}{L} \int_0^{L} f(x,y) \mathrm{d} y $ and dashed 
 quantities represent the magnitude of variations from the $y$-averaged value as a result of turbulent fluctuations. It should be noted that these variations themselves have zero mean.  
 Because the velocity field is homogeneous and isotropic  $\overline{v}=0$. 
 From Eq.~\eqref{eq:fishact} we obtain the equations for $\overline{c}$ and $c^{\prime}$:
\begin{eqnarray}
  \frac{\partial\overline{c}}{\partial t} 
   &=& - \partial_x \overline{\bm{F}} 
           - \mu \overline{H} + \mu \overline{c} (1-\overline{c})+ D {\partial_{xx}} \overline{c}, \label{eq:ap6}  \\        
 \frac{\partial{c'}}{\partial{t}}
    &=& \mu c'  (1-2\overline{c})
                      - \nabla \cdot ( \bm{v} \overline{c}  + \bm{v}c') + \partial_x\overline{\bm{F}} \nonumber \\
                      &&- \mu ({c'}^2 -\overline{H}) + D {\nabla}^2 c'.
\label{eq:ap7}
\end{eqnarray} 
Here $ H = {c'}^2 $ and $\bm{F} = \bm{v}c'$ are, respectively, the autocorrelation and flux of the turbulent fluctuations in the concentration field. We describe their 
time evolution here.  From our numerical simulations, we find that $\overline{H}$ is negligible. 
We further assume:  (a) the turbulence time scales are smaller than the scales associated with the front so that the time variation of 
the turbulent fluctuations in the velocity field can be ignored in the evolution equation for $\bm{F}$ ; (b) isotropic velocity field ${\bm {v v}}= v^2 \mathbb{I}$;  
and (c) the $\tau$ approximation $\overline{{\bm v}{\bm v} \cdot \nabla c'}= \overline{{\bm v} c'}/\tau$ \cite{bran}. Then from Eq. $\eqref{eq:ap7}$ we find that $\overline{\bm F}$ 
boils down to a scalar quantity $\overline{F}$ obeying the following equation: 
\begin{eqnarray}
\frac{\partial \overline{F}}{\partial t} = -v^2\partial_x\overline{c}-\frac{\overline{F}}{\tau_F}.
\label{eq:ap12}
\end{eqnarray}
Here $\frac{1}{\tau_F}=\frac{1}{\tau}-\mu(1-2\overline{c})$, $\tau_F$ is the relaxation time for $\overline{F}$ \cite{bran}, and $\mathbb{I}$ is the 
identity matrix. Assuming that $\overline{F}$ does not vary over the front propagation time scales, using Eq.~\eqref{eq:ap12}, we  get the Fickian form $\overline{F}=-{D}_t\partial_x\overline{c} $, where $D_t= -\tau_{F}v^2$ and $v={v_{rms}}/{\sqrt{2}} $. Because the Fisher front propagates along the horizontal ($x$) direction, the variations along 
the vertical ($y$) direction have been neglected. We, therefore, estimate the eddy diffusivity for our simulations as $D_t= \overline{v_x c'}/(\partial_x \overline{c})$, where $v_x$ is the horizontal component of the velocity. The numerical estimate of $D_t$ for various values of $D$ and $\mu$ are tabulated in Table~\ref{tab}. We find that $D_t$ varies between 
$0.1$ and $0.2$ and is very close with the eddy-diffusivity estimate $D_t=v_0/k_m\approx 0.17$ that we use in the main text.

%%%%%%%%%% Bibliography here %%%%%%%%%%%%%%%


\begin{thebibliography}{10}

\bibitem{wak94}
J. Wakita, K. Komatsu, A. Nakahara,T. Matsuyama and M. Matsushita, J.  Phys.  Soc.  Jpn. {\bf 63},  1205  (1994).

\bibitem{jac94}
E. Ben-Jacob, O. Schochet, A. Tenenbaum, I. Cohen, A. Czirok and T. Viscek, Nature(London) {\bf 368},  46  (1994).

\bibitem{ver08}
N. Verstraeten, K. Braeken, B. Debkumari, M. Fauvart, J. Fransaer, J. Vermant and J. Michiels, Trends Microbiol. {\bf 16},  496  (2008).

\bibitem{swi09}
A. Beer, H.P. Zhang, E.L. Florin, S.M. Payne, E. Ben-Jacob and H.L. Swinney, Proc. Natl. Acad. Sci. U. S. A. {\bf 106},  428  (2009).

\bibitem{kor10}
K.S. Korolev, M. Avlund, O. Hallatschek, and D.R. Nelson, Rev. Mod. Phys. {\bf 82},
   1691  (2010).

\bibitem{den14}
P. Deng, L.d.V. Roditi, D. van Ditmarsch, and J.B. Xavier, New J. Phys. {\bf
  16},  015006  (2014).

\bibitem{kea10}
D.B. Kearns, Nat. Rev. Microbiol. {\bf 8},  634  (2010).

\bibitem{wen}
H.H. Wensink, J. Dunkel, S. Heidenreich, K. Drescher, R.E. Goldstein, H. Löwen and J.M. Yeomans, Proc. Natl. Acad. Sci. U. S. A. {\bf 109},  14308  (2012).

\bibitem{czi00}
A. Czirok, M. Matsushita, and T. Vicsek, Phys. Rev. E {\bf 63},  031915 (2001).

\bibitem{cop09}
M.F. Copeland and D.B. Weibel, Soft Matter {\bf 5},  1174  (2009).

\bibitem{sok07}
A. Sokolov, I.S. Aranson, J.O. Kessler, and R.E. Goldstein, Phys. Rev. Lett. {\bf
  98},  158102  (2007).

\bibitem{sok09}
A. Sokolov, R.E. Goldstein, F.I. Feldchtein, and I.S. Aranson, Phys. Rev. E {\bf 80},
   031903  (2009).

\bibitem{bra15}
V. Bratanov, F. Jenko, and E. Frey, Proc. Natl. Acad. Sci. U. S. A. {\bf 112},  15048
  (2015).


\bibitem{lac99}
A.M. Lacasta, I.R. Cantalapiedra, C.E. Auguet, A. Penaranda and L. Ramirez-Piscina, Phys. Rev. E {\bf 59},  7036  (1999).

\bibitem{gol98}
I. Golding, Y. Kozlovsky, I. Cohen, and E. Ben-Jacob, Phys. A (Amsterdam, Neth.){\bf 260},  510
   (1998).

\bibitem{pig13}
S. Pigolotti, R. Benzi, P. Perlekar, M.H. Jensen, F. Toschi and D.R. Nelson, Theor. Popul. Biol. {\bf 84},  72
  (2013).

\bibitem{bran}
A. Brandenburg, NilsErlandL. Haugen, and N. Babkovskaia, Phys. Rev. E {\bf 83},
  016304  (2011).

\bibitem{bha15}
A.K. Bhattacharjee, K. Balakrishnan, A.L. Garcia, J.B. Bell and A. Donev, J. Chem. Phys. {\bf 142},  224107  (2015).

\bibitem{ben98}
E. Ben-Jacob, I. Cohen, and D.L. Gutnick, Annu. Rev. Microbiol. {\bf 52},  779
  (1998).

\bibitem{czi96}
A. Czirok, E. Ben-Jacob, I. Cohen, and T. Vicsek, Phys. Rev. E {\bf 54},  1791
  (1996).

\bibitem{ben97}
E. Ben-Jacob, Contemp. Phys. {\bf 38},  205  (1997).

\bibitem{fis37}
R.A. Fisher, Ann. Eugenics {\bf 7},  335  (1937).

\bibitem{abr98}
E.R. Abraham, Nature(London) {\bf 391},  577  (1998).

\bibitem{mar03}
A.P. Martin, Prog. Oceanogr. {\bf 57},  125  (2003).

\bibitem{per10}
P. Perlekar, R. Benzi, D.R. Nelson, and F. Toschi, Phys. Rev. Lett. {\bf 105},
  144501  (2010).

\bibitem{ramaswamy1}
S. Ramaswamy, Annu. Rev. Condens. Matter Phys. {\bf 1},  323  (2010).

\bibitem{ber06}
E. Bertin, M. Droz, and G. Gregoire, Phys. Rev. E. {\bf 74},  022101  (2006).

\bibitem{mis10}
S. Mishra, A. Baskaran, and M.C. Marchetti, Phys. Rev. E. {\bf 81},  061916 (2010).

\bibitem{mar13}
M.C. Marchetti, J.F. Joanny, S. Ramaswamy, T.B. Liverpool, J. Prost, M. Rao, and R. Aditi Simha, Rev. Mod. Phys. {\bf 85},  1143  (2013).

\bibitem{ton14}
J. Toner, Y. Tu, and S. Ramaswamy, Ann. Phys.(NY) {\bf 318},  170  (2014).

\bibitem{yan14}
X. Yang, D. Marenduzzo, and M.C. Marchetti, Phys. Rev. E {\bf 89},  012711 (2014).

\bibitem{doo16}
A. Doostmohammadi, S.P. Thampi, and J. Yeomans, Phys. Rev. Lett. {\bf 117},  048102 (2016).

\bibitem{per11}
P. Perlekar, S.S. Ray, D. Mitra, and R. Pandit, Phys. Rev. Lett. {\bf 106}, 054501  (2011).

\bibitem{murray}
J.D. Murray, {\em Mathematical Biology: I. An Introduction, Interdisciplinary Applied Mathematics} (Springer New York,2011).

\bibitem{kol37}
A. Kolmogorov, I. Petrovsky, and N. Piscounov, Moscow University Bulletin of Mathematics {\bf 1}, 1 (1937).

\bibitem{gast1}
M.T. Gastner, B. Oborny, D.K. Zimmermann, and G. Pruessner, Am. Nat. {\bf 174},  E23  (2009).

\bibitem{sap85}
B. Sapoval, M. Rosso, and J.F. Gouyet, J. Phys. Lett. {\bf 46},  L149
  (1985).

\bibitem{lem}
A. Lemarchand, I. Nainville, and M. Mareschal, Europhys. Lett. {\bf 36},
  227  (1996).

\bibitem{bat59}
G.K. Batchelor, J. Fluid Mech. {\bf 5},  113  (1959).

\bibitem{kra68}
R.H. Kraichnan, Phys. Fluids {\bf 11},  945  (1968).

\end{thebibliography}
\end{document}